\documentclass[a4paper,11pt]{article}
\usepackage{pos}

\usepackage{pennames}
\usepackage{xspace}
\usepackage{units}
\usepackage{multirow}

\manuallySeparateAuthors

\def\dime{\textsc{Dime}}
\def\prof{{\sc Professor}}

\def\d{{\mathrm d}}
\def\P{{\rm I\!P}}

\newcommand{\GeV}{\ensuremath{\,\text{Ge\hspace{-.08em}V}}\xspace}

\newcommand{\pt}{\ensuremath{p_{\mathrm{T}}}\xspace}

\title{Central exclusive production in CMS+TOTEM}

\author*[a]{Ferenc Sikl\'er}
\author[]{\\for the CMS and TOTEM Collaborations}

\affiliation[a]{Wigner Research Centre for Physics,\\
  Konkoly-Thege M. \'ut 29-33, 1121 Budapest, Hungary}

\emailAdd{sikler.ferenc@wigner.hu}

\abstract{
The central exclusive production of charged hadron pairs in pp collisions at a
centre-of-mass energy of 13 TeV is examined, based on data collected in a
special high-$\beta^*$ run of the LHC. Events are selected by requiring both
scattered protons detected in the TOTEM Roman pots, exactly two oppositely
charged identified particles in the CMS silicon tracker, and the
energy-momentum balance of these four particles. The nonresonant continuum
processes are studied with the invariant mass of the centrally produced
two-pion system in the resonance-free region, $m < 0.7~\mathrm{GeV}$ or $m >
1.8~\mathrm{GeV}$. Differential cross sections as functions of the azimuthal
angle between the surviving protons, squared four-momenta, and two-hadron
invariant mass are measured in a wide region of scattered proton transverse
momenta $0.2~\mathrm{GeV} < p_\text{1,T}, p_\text{2,T} < 0.8~\mathrm{GeV}$ and
for hadron rapidities $|y| < 2$. A rich structure of interactions related to
double pomeron exchange emerges. The parabolic minimum in the distribution of
the two-proton azimuthal angle is observed for the first time. It can be
understood as an effect of additional pomeron exchanges between the protons
from the interference between the bare and the rescattered amplitudes. After
model tuning, various physical quantities related to the pomeron cross section,
proton-pomeron and hadron-pomeron form factors, trajectory slopes and
intercepts, as well as coefficients of diffractive eigenstates of the proton
are determined.
}

\FullConference{The European Physical Society Conference on High Energy Physics (EPS-HEP2023)\\
 21-25 August 2023\\
Hamburg, Germany\\}

\begin{document}
\maketitle

\section{Introduction}

In collisions of protons, the exclusive central production of a few particles
offers a clean laboratory for the study of various
phenomena~\cite{Albrow:2010yb}. At high energies, for not too small momentum
transfers, these processes are dominated by double pomeron exchange. They
provide a gluon rich environment which may lead to the creation of hadrons free
of valence quarks, the glueballs.

\begin{figure}[t]

 \centering
 \includegraphics[width=\textwidth]{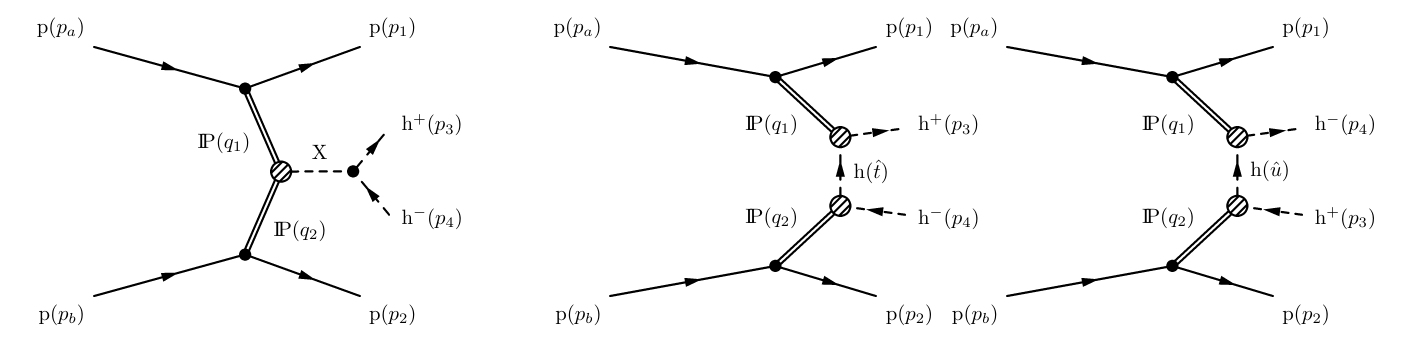}

 \caption{Born-level Feynman diagrams for central exclusive production of
hadron pairs via double pomeron exchange, depicting resonant (left) and
nonresonant continuum (rightmost two) contributions.}

 \label{fig:feynman}

\end{figure}

Born-level Feynman diagrams of processes playing roles in central exclusive
production of charged hadron pairs via double pomeron exchange are shown in
Fig.~\ref{fig:feynman}. We distinguish two main processes. The two pomerons may
fuse through a short-lived resonance (X), which in turn decays to a pair of
oppositely charged hadrons. Alternatively, the two pomerons may interact via
the exchange of a virtual hadron (h), leading to the production of a pair of
oppositely charged hadrons in a nonresonant process, the topic of the present
study. Both $t$-channel and $u$-channel graphs should be taken into account.

Such a simple picture gets complicated when the various effects of suppression,
absorption, and other corrections are properly taken into account. Additional
pomeron exchanges between the incoming (and outgoing) protons cause
interference between the bare and the rescattered amplitudes. The resulting
quantity is often referred to as the eikonal survival factor. The interference
is expected to lead to interesting diffractive dip phenomena in the angular
distributions~\cite{Harland-Lang:2013dia}.

\section{Data taking and analysis}

A detailed description of the CMS detector, together with a definition of the
relevant kinematic variables, can be found in Ref.~\cite{Chatrchyan:2008zzk}.
The TOTEM detector is described in Ref.~\cite{Anelli:2008zza}.
The data were taken in a special $\beta^* = 90\,\mathrm{m}$ run of the LHC, in
2018 ($\beta^*$ is the value of the amplitude function at the collision point).
With such a high $\beta^*$ setting the beam divergence is reduced, hence the
forward detectors are able to probe the elastic-scattering regime at small
scattering angles and small transverse momenta.

The Roman pots (RPs) measure the direction of the scattered proton. Its
transverse momentum is inferred on the assumption that the total momentum of
the proton has not changed in the collision. The acceptance of the stations is
not azimuthally uniform~\cite{Anelli:2008zza} and the coverage in the
transverse momentum of the scattered protons is $0.175\GeV < |p_{1/2,y}| <
0.670\GeV$. The acceptance maps of the two arms are correlated since signals
from both arms are used for triggering. In addition, their efficiency depends
on the individual silicon strip detector efficiencies, which change with time.
The silicon tracker covers the region $|\eta| < 3$. This translates into
acceptance for centrally produced hadrons for $|y| < y_\text{max}$, where
$y_\text{max} \approx 2.0$ in the case of $\Pgpp\Pgpm$, and $y_\text{max}
\approx 1.6$ for $\PKp\PKm$ and $\Pp\Pap$. Tracking is efficient for $\pt >
0.1\GeV$, but the particle identification capabilities are substantially
reduced at higher momenta.

For this data set the Level 1 (hardware) trigger requires detected protons in
each RP arm, in various configurations. Only the data from double pomeron
exchange triggers are used in the analysis, not those from the elastic trigger.
The high-level trigger (HLT) has multiple components: the pixel activity
and track filters.
 
The data are studied as functions of the four-momentum transfers squared
$(t_1,t_2)$ or equivalently $(p_\text{1,T},p_\text{2,T})$, and the azimuthal
angle $\phi$ between the momentum vectors of the two scattered protons in the
transverse plane.
The measurement is fully corrected without using Monte Carlo simulations,
except those describing low-energy phenomena needed for tracking efficiency
correction.
The results are corrected for the momentum acceptance and the effect of the
elastic trigger veto of the RPs using data, for
the local straight track reconstruction efficiency in the RPs, based
on the hit structure of each tracklet at the strip level.
Data are corrected for trigger, reconstruction, and particle identification
efficiencies of the charged hadron pair in the silicon tracker.

The classification of events is largely based on momentum conservation in the
transverse plane. Once corrected for the effects of the beam crossing angle,
the system of the incoming protons has zero momentum. Therefore, the sum of the
momenta of the scattered protons and the other particles created in the
collision should also be zero.
The momentum conservation in the longitudinal direction is already utilised for
the calculation of longitudinal momenta of the scattered protons, since the
resolution from a direct measurement would be poor.

\begin{figure}

 \centering
 \includegraphics[width=0.75\textwidth]{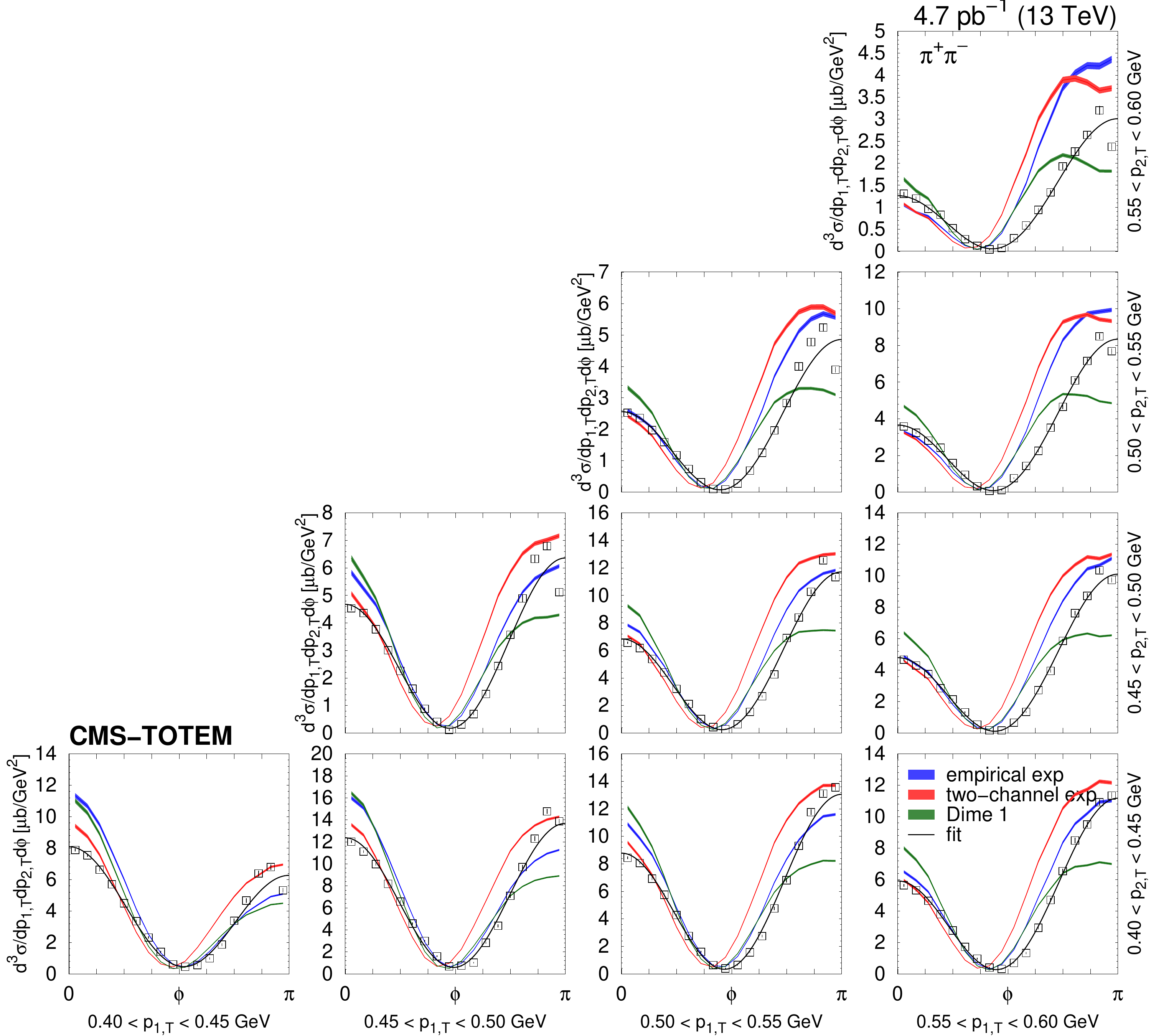}

 \vspace{0.00\textheight}
 \includegraphics[width=0.75\textwidth]{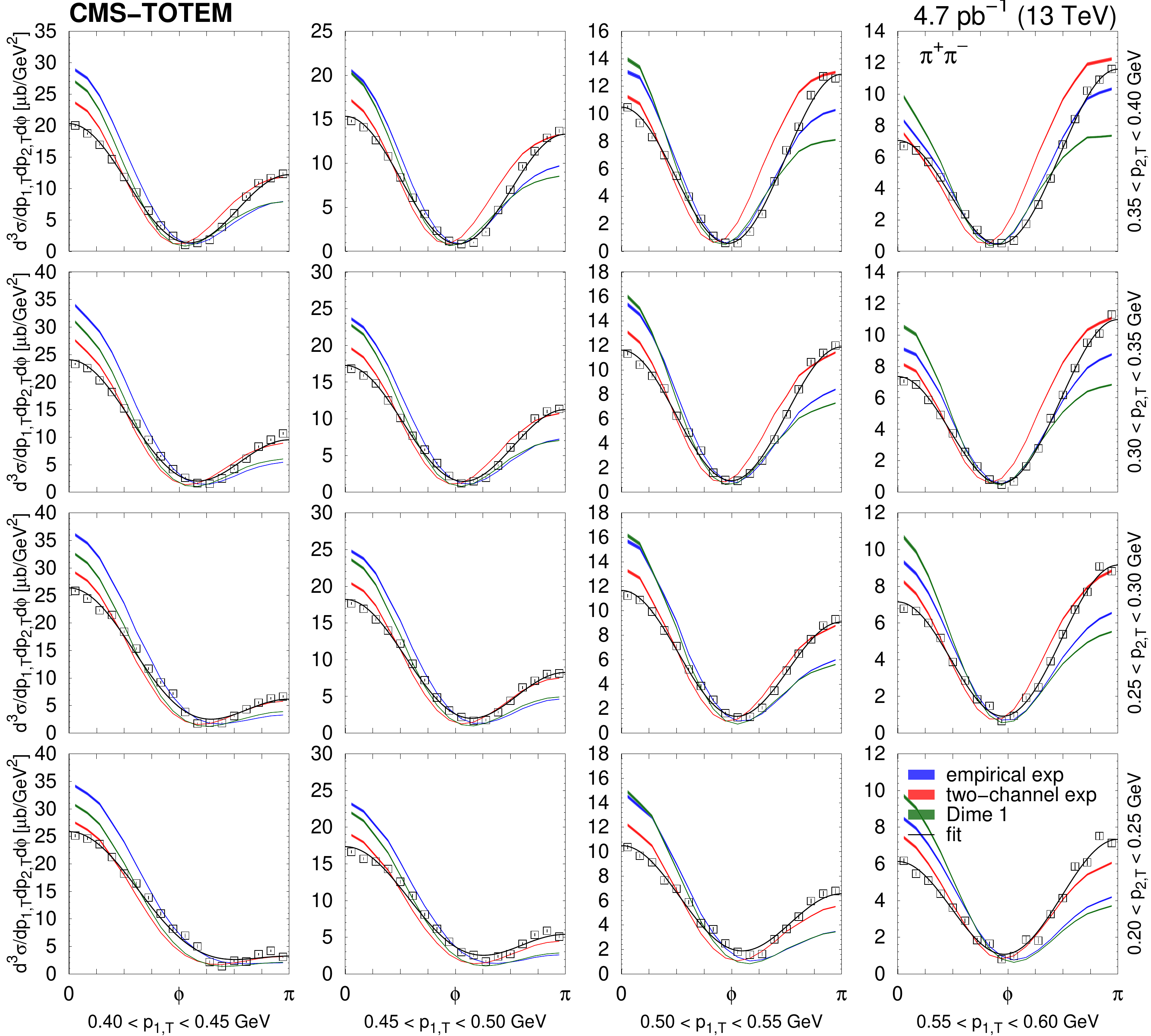}

 \caption{Distribution of $\d^3\sigma/\d p_\text{1,T} \d p_\text{2,T} \d\phi$ as a
function of $\phi$ in several $(p_\text{1,T},p_\text{2,T})$ bins, in units of
$\mu\mathrm{b}/\GeV^2$. Measured values (black symbols) are shown together with
the predictions of the empirical and the two-channel models (coloured symbols)
using the tuned parameters for the exponential proton-pomeron form factors (see
text for details). Curves corresponding to {\dime} (model 1) are also plotted.
Results of fits with the form $[A(R - \cos\phi)]^2 + c^2$ are plotted with
curves. The error bars indicate the statistical uncertainties.}

 \label{fig:dndphi_withmc}

\end{figure}

\section{Results}

The measured distributions are the following:
 distribution of the azimuthal angle $\phi$ between the
       scattered proton momenta,
       $\d^3\sigma/\d p_\text{1,T}\,\d p_\text{2,T}\,\d\phi$;
 distribution of the two-hadron invariant mass $m$,
       $\d^3\sigma/\d p_\text{1,T}\,\d p_\text{2,T}\,\d m$;
 distribution of the squared four-momentum $\max(\hat{t},\hat{u})$ of
       the virtual meson,
       $\d^3\sigma/\d p_\text{1,T}\,\d p_\text{2,T}\,\d \max(\hat{t},\hat{u})$,
in the range $0.2 < p_\text{1,T},p_\text{2,T} < 0.8\GeV$.

We focus on $\Pgpp\Pgpm$ production because this system has a wide invariant
mass window ($0.35 < m < 0.65\GeV$) without resonance contributions, contrary
to the $\PKp\PKm$ case.
A selection of distributions of $\d^3\sigma/\d p_\text{1,T} \d p_\text{2,T}
\d\phi$ as a function of $\phi$ in several $(p_\text{1,T},p_\text{2,T})$ bins
are shown in Fig.~\ref{fig:dndphi_withmc}. The differential cross sections are
given in units of $\mu\mathrm{b}/\GeV^2$.
Except for the lowest transverse momentum bins, they feature a minimum where
the differential cross section gets close to zero, while local maxima at $\phi
= 0$ and $\pi$ are also present. The distributions are asymmetric at low and
high transverse momenta, but there exists a symmetric region around
$p_\text{1,T} + p_\text{2,T} \approx 0.8-0.9\GeV$.

The data can be fitted with a simple functional form
\begin{equation}
 \frac{d^3\sigma}{\d p_\text{1,T} \d p_\text{2,T} \d\phi} = [A(R - \cos\phi)]^2 + c^2,
 \label{eq:ARc}
\end{equation}

\noindent
where $A$, $R$, and $c$ are functions of $(p_\text{1,T},p_\text{2,T})$.
If the total amplitude crosses zero at a given $\phi$, its squared value will
have a parabolic minimum. Such a dip at $\phi = \arccos R$ can be understood as
an effect of additional pomeron exchanges between the incoming protons,
resulting from the interference between the bare and the rescattered
amplitudes~\cite{Harland-Lang:2013dia}. The term containing $c$ is added
incoherently, it is small and is present to improve the quality of the fit.

The dependences of the parameters $A$, $R$, and $c$ on $(t_1,t_2)$ can be well
described though
\begin{align}
 A(t_1,t_2) &= 4 \sqrt{t_1 t_2} \cdot A_0 e^{b(t_1 + t_2)}, &
 c(t_1,t_2) &= c_0 e^{d(t_1 + t_2)},
 \\
 R(t_1,t_2) &\approx
  \frac{1.2(\sqrt{-t_1} + \sqrt{-t_2}) - 1.6 \sqrt{t_1 t_2} - 0.8}
       {\sqrt{t_1 t_2} + 0.1}.
\end{align}

\noindent
While the parametrisation overall gives a good description of the data, there
are some deviations at low and high $-(t_1+t_2)$ values for $A$ and $c$,
respectively. The fitted values are $A_0 = 10.6 \pm 0.2
\sqrt{\mathrm{nb}}/\GeV^3$, $b = 3.9 \pm 0.1 \GeV^{-2}$, while $c_0 = 2.1 \pm
0.1 \sqrt{\mathrm{nb}}/\GeV$, $d = 3.8 \pm 0.1 \GeV^{-2}$.

\section{Model tuning}

In the following we will deal with three models (empirical, one-channel, and
two-channel), and three parametrisations (exponential, Orear-type, and
power-law) of the proton-pomeron form factor.
The full list of model parameters can be found in
Ref.~\cite{CMSCollaborationTOTEM:2023dbs}.
In the case of the empirical model the rescattering amplitude comes from a
simple parametrisation of the elastic differential proton-proton amplitude,
fitted to the measured cross sections. The one-channel model assumes ground
state protons (one eigenstate), while the two-channel model works with two
diffractive proton eigenstates.
{\dime}~\cite{Harland-Lang:2013dia} (v1.07) is a MC event generator for
exclusive meson pair production via double pomeron exchange.

For the tuning of physics parameters the tool {\prof}~\cite{Buckley:2009bj}
is employed.
It parametrises the per-bin generator response to
parameter variations and numerically optimises the parameterised behaviour.
The $\chi^2/\text{dof}$ values are in the range $1.6-2.0$ (empirical),
$1.2-1.5$ (one-channel), and $1.0-1.3$ (two-channel). Good fits are achieved
with the one-channel or the two-channel models using exponential or Orear-type
form factors, while the numerically best one is the two-channel model with
exponential parametrisation of the proton-pomeron form factor.
It is clear that the power-law parametrisation of the proton-pomeron
form-factor is disfavoured by our data. The empirical model also has a high
goodness-of-fit value, although with fairly few parameters.

\begin{figure}[t]

 \centering
 \includegraphics[height=0.4\textheight]{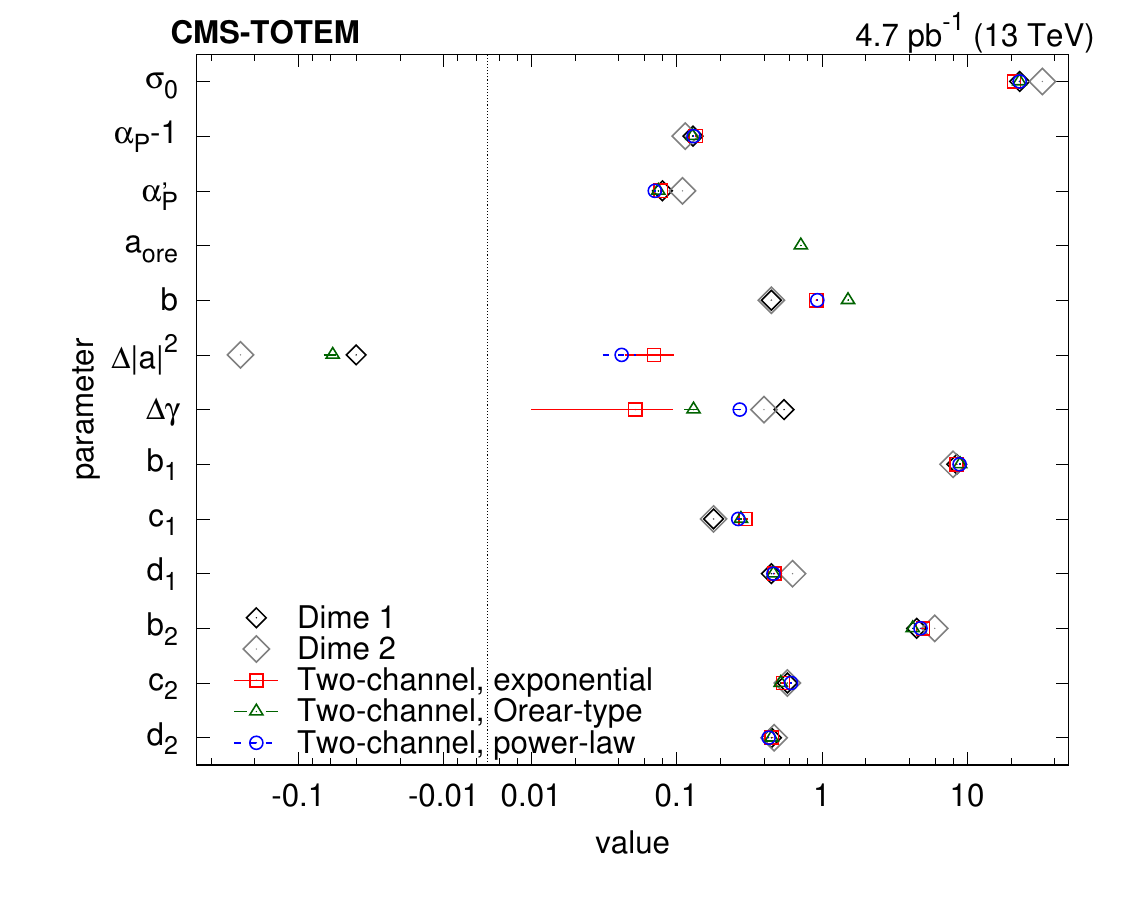}

 \caption{Values of best parameters for the two-channel model with several
choices of the proton-pomeron form factor (exponential, Orear-type, power-law).
Parameter values of models describing the elastic differential proton-proton
cross section from Ref.~\cite{Khoze:2013dha} are also indicated ({\dime} models
1 and 2).}

 \label{fig:tune_results}

\end{figure}

The values of the fitted parameters are shown in Fig.~\ref{fig:tune_results}.
In the case of the two-channel model, parameter
values of two models describing the elastic differential proton-proton cross
section from Ref.~\cite{Khoze:2013dha} are also indicated ({\dime}~1 and 2).
Settings of {\dime}~1 agree well with the best tuned values of the parameters,
with the exception of $\Delta|a|^2$ and $\Delta\gamma$.
The extracted values of $\sigma_0$ are mostly stable, except for the
one-channel exponential case, and disagree with early $C_\P$ estimates.

The distributions of $\d^3\sigma/\d p_\text{1,T} \d p_\text{2,T} \d\phi$ in the
nonresonant region ($0.35 < m < 0.65\GeV$) as a function of $\phi$ in several
$(p_\text{1,T},p_\text{2,T})$ bins are shown in Fig.~\ref{fig:dndphi_withmc}.
Measured values are shown together with the predictions of the empirical and
the two-channel models using the tuned parameters for the exponential
proton-pomeron form factors.
Curves corresponding to {\dime} (model 1) are also plotted: this generator
gives a poor description of the data for $\phi > \pi/2$. While the various
tuned models give a much better description, there are some regions with
sizeable disagreements pointing to the need for further theoretical
developments, specially for low $p_\text{1,T}$ and low $p_\text{2,T}$ as well
as high $p_\text{1,T}$ and high $p_\text{2,T}$ combinations.

\noindent 
Further details and all results of the analysis can be found in
Ref.~\cite{CMSCollaborationTOTEM:2023dbs}.
This work was partially supported by the National Research, Development and
Innovation Office of Hungary (K~128786).

\bibliographystyle{JHEP}
\bibliography{sikler_cex_hep-eps-2023}

\providecommand{\href}[2]{#2}\begingroup\raggedright\begin{thebibliography}{1}

\bibitem{Albrow:2010yb}
M.~Albrow, T.~Coughlin and J.~Forshaw, \emph{{Central Exclusive Particle
  Production at High Energy Hadron Colliders}},
  \href{https://doi.org/10.1016/j.ppnp.2010.06.001}{\emph{Prog. Part. Nucl.
  Phys.} {\bfseries 65} (2010) 149}
  [\href{https://arxiv.org/abs/1006.1289}{{\ttfamily 1006.1289}}].

\bibitem{Harland-Lang:2013dia}
L.A.~Harland-Lang, V.A.~Khoze and M.G.~Ryskin, \emph{{Modelling exclusive meson
  pair production at hadron colliders}},
  \href{https://doi.org/10.1140/epjc/s10052-014-2848-9}{\emph{Eur. Phys. J. C}
  {\bfseries 74} (2014) 2848}
  [\href{https://arxiv.org/abs/1312.4553}{{\ttfamily 1312.4553}}].

\bibitem{Chatrchyan:2008zzk}
{\scshape CMS} collaboration, \emph{The {CMS} experiment at the {CERN} {LHC}},
  \href{https://doi.org/10.1088/1748-0221/3/08/S08004}{\emph{JINST} {\bfseries
  3} (2008) S08004}.

\bibitem{Anelli:2008zza}
{\scshape TOTEM} collaboration, \emph{{The TOTEM experiment at the CERN Large
  Hadron Collider}},
  \href{https://doi.org/10.1088/1748-0221/3/08/S08007}{\emph{JINST} {\bfseries
  3} (2008) S08007}.

\bibitem{CMSCollaborationTOTEM:2023dbs}
{CMS and TOTEM collaborations}, \emph{{Nonresonant central exclusive production
  of charged hadron pairs in proton-proton collisions at $\sqrt{s} =
  13~\mathrm{TeV}$}},
  \href{http://cds.cern.ch/record/2867988}{CMS-PAS-SMP-21-004} (2023).

\bibitem{Buckley:2009bj}
A.~Buckley, H.~Hoeth, H.~Lacker, H.~Schulz and J.E.~von Seggern,
  \emph{{Systematic event generator tuning for the LHC}},
  \href{https://doi.org/10.1140/epjc/s10052-009-1196-7}{\emph{Eur. Phys. J. C}
  {\bfseries 65} (2010) 331} [\href{https://arxiv.org/abs/0907.2973}{{\ttfamily
  0907.2973}}].

\bibitem{Khoze:2013dha}
V.~Khoze, A.~Martin and M.~Ryskin, \emph{{Diffraction at the LHC}},
  \href{https://doi.org/10.1140/epjc/s10052-013-2503-x}{\emph{Eur. Phys. J. C}
  {\bfseries 73} (2013) 2503}
  [\href{https://arxiv.org/abs/1306.2149}{{\ttfamily 1306.2149}}].

\end{thebibliography}\endgroup

\end{document}